\documentclass[ reprint, amsmath,amssymb, prl, superscriptaddress, aps]{revtex4-2}
\usepackage{graphicx}% Include figure files
\usepackage{dcolumn}% Align table columns on decimal point
\usepackage{bm}% bold math
\usepackage{xcolor}
\usepackage{amsmath}
\usepackage{soul}
\usepackage{comment}
\usepackage{makecell}
\newcommand{\VTNCE}{V[TCNE]$_x$}
 
\newcommand{\icite}{~\hl{[cite(s)]}}

\begin{document}

\preprint{APS/123-QED}

\title{Cooling and Squeezing a Microwave Cavity State with Magnons Using a Beam Splitter Interaction}% Force line breaks with \\

\author{Qin Xu}
\affiliation{Department of Physics, Cornell University, Ithaca NY 14853}
\author{Gregory D. Fuchs}
\affiliation{School of Applied and Engineering Physics, Cornell University, Ithaca NY 14853}

\date{\today}% It is always \today, today,

\begin{abstract}
We propose two geometries to realize a significant beam splitter interaction (XZ coupling) between magnons and a 2D microwave cavity mode. In both setups the cavity is analogous to the mechanical oscillator in a conventional optomechanical setup. The backaction effects are calculated with realistic experimental parameters. The analytical results show that we can not only make the backaction damping (anti-damping) rate larger than the bare microwave resonator damping rate, but that we can also achieve quantum squeezing of the resonator where the uncertainty in one quadruture (charge or current) is smaller than its zero point fluctuation.  

\end{abstract}

\maketitle

%\tableofcontents

%\section{Introduction}

Cavity magnonics has emerged as a promising building block of hybrid quantum systems. For example, magnons can enhance the coupling between microwave cavities and long-lived quantum spins like color centers~\cite{candido2020predicted,lee2020nanoscale,mccullian2020broadband}. Also, because of the nonreciprocal nature of magnetism, cavity magnonics have been proposed to create nonreciprocal entanglement in hybrid quantum systems \cite{chakraborty2023nonreciprocal, chen2023nonreciprocal, zheng2024nonreciprocal}. The linear oscillator coupling (XX coupling) between the cavity and magnons has been studied both theoretically and experimentally~\cite{Soykal2010,Huebl2013}. Specifically, the 2D patterning of both a superconducting (SC) resonator and a magnetic microstructure has made the coupled system scalable while preserving strong XX coupling \cite{guo2023strong, BhoiB2014Sopc, haygood2021strong, Li2019, Stenning:13, xu2024strong}. For example, recent studies have shown that the lowest damping magnetic materials, such as ytrrium iron garnet (YIG) \cite{guo2023strong, devitt2024distributed} and vandium tetracyanoethylene (\VTNCE)~\cite{Franson2019, zhu2020organic}, can be patterned on the micron scale on 2D structures.
%For example, recent studies have tried to pattern the lowest damping magnetic materials, such as ytrrium iron garnet (YIG) \cite{guo2023strong, devitt2024distributed} and vandium tetracyanoethylene (\VTNCE)~\cite{Franson2019, zhu2020organic}, while preserving the low damping.

%Cavity magnonics is currently a popular area of research. The linear oscillator coupling (XX coupling) between the cavity and magnons has been studied both theoretically and experimentally \icite. Specifically, the 2D patterning of both the SC resonator (cavity) and magnetic materials has made the coupled system scalable while preserving strong XX coupling \icite. For example, recent studies have shown that the lowest damping magnetic materials, such as ytrrium iron garnet (YIG) \icite and vandium tetracyanoethylene (\VTNCE)~\icite, can be patterned on the micron scale on 2D structures. Magnets in the cavity can play the role of couplers in hybrid quantum systems. For example, they can enhance the coupling between microwave cavities and long lived quantum spins like color centers \icite. Also, because of magnon’s nonreciprocal property, cavity magnonics has been proposed to create nonreciprocal entanglement in the hybrid quantum systems \cite{chakraborty2023nonreciprocal, chen2023nonreciprocal, zheng2024nonreciprocal}.

The beam splitter interaction (XZ coupling) between microwave (MW) cavities and magnons has not yet been fully analyzed. Such nonlinear coupling has been analyzed and experimentally studied in other platforms including an optical cavity coupled to a mechanical oscillator \cite{dorsel1983optical}, a microwave cavity coupled to a mechanical oscillator \cite{regal2008measuring, teufel2011circuit}, and magnons coupled to a mechanical oscillator \cite{wang1970spin, zhang2016cavity}. These systems can be used for sideband cooling, antidamping, squeezing of the mechanical mode \cite{mancini1998optomechanical, cohadon1999cooling, kronwald2013arbitrarily, wollman2015quantum, pirkkalainen2015squeezing, youssefi2023squeezed}, quantum non-demolition detection of the mechanical mode \cite{lecocq2015quantum, clerk2008back, hertzberg2010back, suh2014mechanically}, optomechanically induced transparency \cite{schliesser2009cavity, agarwal2010electromagnetically, weis2010optomechanically, safavi2011electromagnetically}, and storing information in the long lived mechanical degrees of freedom \cite{chang2011slowing} to name a few applications.

In this letter, we propose two geometries to realize the XZ interaction between magnons and MW cavity that we refer to as ``top coplanar waveguide (CPW) setup" with YIG as the magnon host material and ``45 degree alignment setup" with \VTNCE ~as the magnon host material. The 45 degree alignment setup also uses XX coupling between magnons and MW photons to drive the magnon sidebands more efficiently \cite{shen2023observation}, which limits the MW power at the sample. In both setups the MW cavity fundamental mode is analogous to the mechanical oscillator in a conventional optomechanical system. Comparing to previous optomagnonic systems \cite{liu2016optomagnonics, viola2016coupled}, our system also has the XZ coupling between photons and magnons, but with the photon frequency much smaller than the magnon frequency. This is advantageous because the SC resonator has a much lower damping rate than magnons in general. Our analytical results show that, with realistic experimental parameters, we can both make the backaction damping (anti-damping) rate larger than the bare MW resonator damping rate and achieve quantum squeezing of the MW resonator where the fluctuation of one quadrature (charge or current) is below its zero point fluctuation (ZPF) \cite{wollman2015quantum, kim1989properties}.

There are several advantages for cooling and squeezing MW resonators with magnons. Firstly, comparing to mechanical oscillators, MW resonators are easier to fabricate, can be more readily measured with a waveguide, and are easier to couple to other MW devices like SC qubits. Also, they have a higher frequency than most mechanical oscillators, which makes it easier to work in the quantum regime. Secondly, the magnon frequency can be easily tuned with external magnetic field, making it straightforward to explore many aspects of the system. Also, their stray field and nonreciprocal properties can be used as we build toward a larger hybrid quantum system, where the magnons are XZ coupled to the resonator while also XX coupled to other systems like long-lived spins.

%\section{Proposed experimental setup}
%\label{prop_setup}

\begin{figure}
\includegraphics[width = \columnwidth]{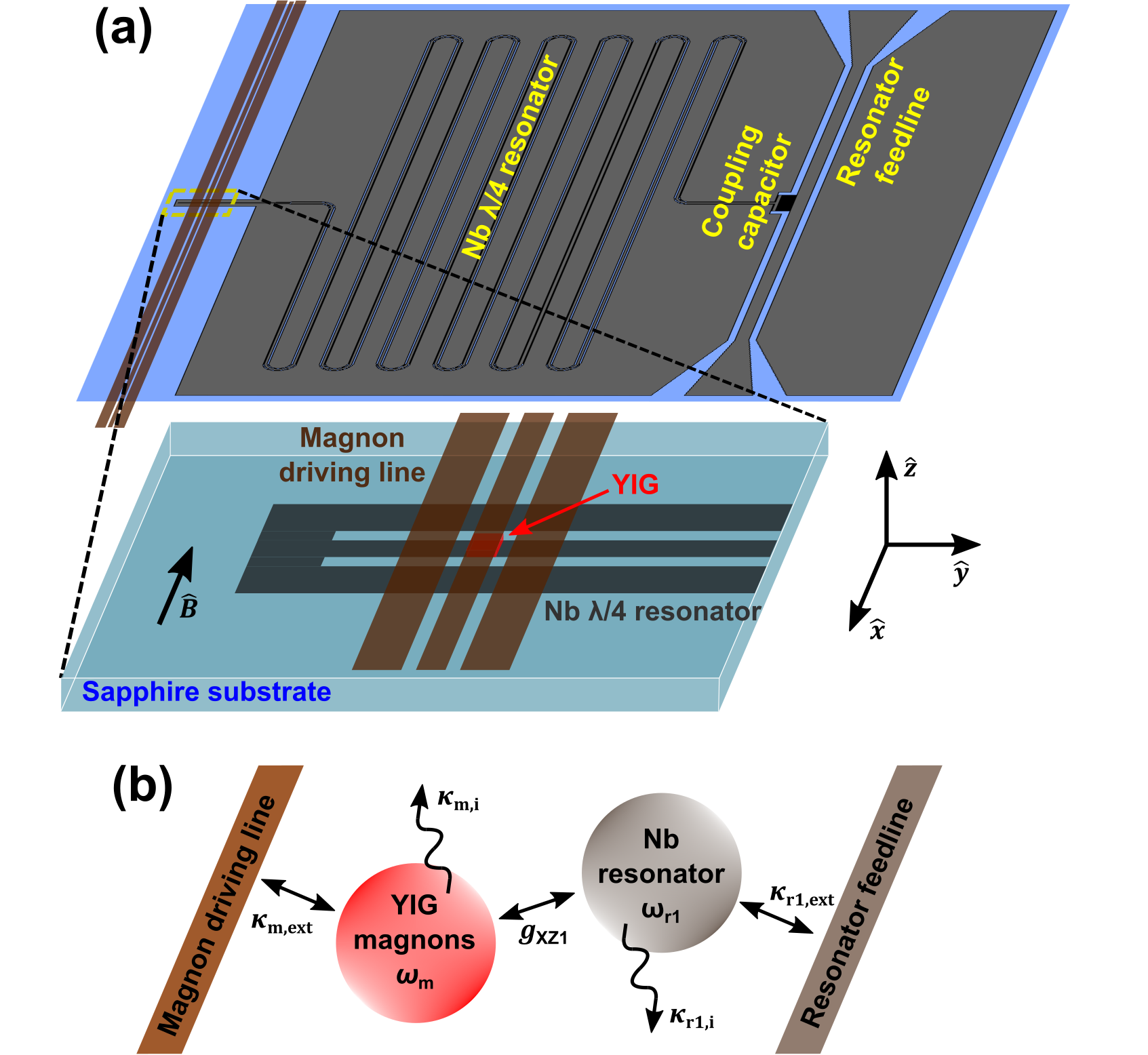}
\caption{(a) Schematic plot of the top CPW setup and a zoomed in plot at the current antinode of the Nb $\lambda/4$ resonator. The waveguide at the current antinode is along the $\hat{y}$ direction, which is perpendicular to the external in-plane static field $\hat{B} = -B\hat{x}$. A  YIG microstructure is on the resonator's central wire and is near the current antinode of the resonator. The resonator can be driven and detected by a resonator feedline with an external coupling rate $\kappa_{r1,ext}$. The magnons in YIG are driven by a magnon driving line from above with coupling rate $\kappa_{m,ext}$. The magnon driving line is along the $\hat{x}$ direction right above YIG. With this geometry there is only XZ coupling between magnons and the Nb resonator (the last term in equation \ref{Hamil_1}) and the MW power from the magnon driving line will perpendicularly pump the magnons in YIG. The crosstalk between the magnon driving line and the resonator is negligible. The YIG's dimension is 5 $\mu$m $ \times $ 5 $\mu$m $ \times$ 1 $\mu$m and the width of the central wire where YIG is placed on is $w=5$ $\mu$m. (b) Scheme of all the dampings and couplings in the top CPW setup. The total damping rate of magnons (resonator) is the sum of the internal and external damping rate: $\kappa_m = \kappa_{m,i} + \kappa_{m,ext}$ ($\kappa_{r1} = \kappa_{r1,i} + \kappa_{r1,ext}$).}
\label{top_CPW_geometry_pic}
\end{figure}

The proposed experimental geometry is shown in Fig.~\ref{top_CPW_geometry_pic}, consisting of a niobium film patterned into a $\lambda/4$ resonator that is capacitively coupled to a resonator feedline that can excite and probe the resonator. At the shorted end of the resonator, a micropatterned magnet with low damping (from this geometry we will assume YIG) is positioned on the central conductor of the resonator.  Fabricated above and crossing the micromagnet and the $\lambda/4$ resonator is a Nb magnon driving line that can drive ferromagnetic resonance in the micromagnet.  For this setup, the external magnetic field $\hat{B}$ will be oriented in the plane along the central conductor of the magnon driving line.  Since the magnon driving line and the resonator's inductor wire meet at the current antinode of the resonator perpendicularly, there is neither capacitive nor inductive coupling between them to leading order. The impedance of the $\lambda/4$ resonator is $Z_0 = 50\, \Omega$, same as the impedance of the CPW.

%the magnon driving line is for driving YIG magnons which is XZ coupled to the 2D $\lambda/4$ resonator, and the resonator can be excited and probed by another resonator feedline. The external in-plane static field $\hat B$ is perpendicular to the resonator inductor wire at the current antinode. With this geometry the oscillating AC field from the inductor wire of the resonator will add up to the external field $\hat B$, which will modulate the Larmor frequency of the magnons. Since the magnon driving line and the resonator's inductor wire meet at the current antinode of the resonator perpendicularly, there is neither capacitive nor inductive coupling between them. The impedance of the $\lambda/4$ resonator is $Z_0 = 50\, \Omega$, same as the impedance of the CPW. The resonator and all the CPW's can be made from superconducting Niobium.

The Hamiltonian of this system is:
\begin{equation}
    {\cal H}/\hbar = \omega_{r1} \hat{a}^\dagger_{r1} \hat{a}_{r1}+\omega_{m}(B)\hat{b}^{\dagger}\hat{b} - g_{XZ1}\hat{b}^{\dagger}\hat{b}\left(\hat{a}_{r1}+\hat{a}^\dagger_{r1}\right)
    \label{Hamil_1}
\end{equation}
where the first 2 terms are the energies of the superconducting resonator's fundamental mode (denoted by subscript $r1$) and the Kittel mode of the magnet (subscript $m$), respectively. The third term is the XZ coupling between the resonator's fundamental mode and the Kittel mode with coupling strength $g_{XZ1} = G\Phi_{ZPF}$, where $\Phi_{ZPF} \propto I_{ZPF}$ is the zero point fluctuation of fundamental mode flux and $G$ is the magnon frequency shift per unit flux.

In a conventional optomechanical system, a position change of the mechanical oscillator will cause a frequency shift of the photon cavity photon, and $g_{{XZ}}$ is equal to the cavity frequency shift caused by the zero point fluctuation of the mechanical oscillation position $x_{ZPF}$ \cite{aspelmeyer2014cavity}. In our system, $g_{XZ1}$ is equal to the Kittel mode frequency shift caused by the zero point fluctuation of the resonator's flux, $\Phi_{ZPF}$:
\begin{equation*}
\begin{split}
    g_{XZ1} &= \frac{\partial \omega_{m}}{\partial \Phi} \Phi_{ZPF}
    = \frac{\partial \omega_{m}}{\partial I} I_{ZPF} = \frac{\partial \omega_{m}}{\partial I} \sqrt{\frac{2\hbar}{\pi Z_0}}\omega_{r1} \\
    &= \gamma \frac{\mu_0}{2w} \sqrt{\frac{2\hbar}{\pi Z_0}}\omega_{r1}
    = 2\pi \times 12.835\, \text{Hz}
\end{split}
\end{equation*}
where $I_{ZPF} = \sqrt{\frac{2\hbar}{\pi Z_0}}\omega_{r1}$ is the ZPF of current at the current antinode of the $\lambda/4$ resonator (see the SI for details \cite{SIref}). Note that $g_{XZ1}$ can be increased by decreasing the resonator impedance. This can be done by decreasing the gap to central wire ratio of the $\lambda/4$ CPW, or by changing the $\lambda/4$ resonator into a lumped element resonator with a much smaller impedance \cite{mckenzie2019low}.

In the classical regime, we can denote the magnon amplitude $\beta_m(t) = \langle \hat b(t) \rangle$, the resonator amplitude $\alpha_{r1}(t) = \langle \hat a_{r1}(t) \rangle$. Also, we denote the flux of the resonator as $\Phi(t) = \langle \hat \Phi(t) \rangle = \Phi_{ZPF}(\langle\hat a_{r1}(t) + \hat a_{r1}^{\dagger}(t)\rangle)$. We define $L_1 = \sqrt{\frac{Z_0}{\omega_{r1}}}$ and $C_1 = \sqrt{\frac{1}{\omega_{r1} Z_0}}$. Similar to an optomechanical system \cite{aspelmeyer2014cavity}, the equations of motion for an XZ coupled system are:
\begin{align*}
    \dot \beta_m &= -\frac{\kappa}{2} \beta_m + i(\Delta + G\Phi) \beta_m 
    \\
    C_1 \ddot \Phi &= -C_1 \omega_{r1}^2 \Phi - C_1 \kappa_{r1} \dot \Phi + \hbar G \left| \beta_m \right|^2
\end{align*}
where we have used the rotating frame of the external drive frequency $\omega_d$ and $\Delta = \omega_d - \omega_m$. In this case $C_1$ and $\Phi$ are analogous to the effective mass and the position of the mechanical oscillator, respectively, in conventional optomechanics.

After linearizing and analyzing in the frequency domain, we find that the backaction will shift the CPW resonator frequency by \cite{SIref}
\begin{widetext}
\begin{eqnarray*}
    \delta \omega_{r1} = n_m g_{XZ1}^2 \frac{\omega_{r1}}{\omega} \left(\frac{\Delta + \omega}{(\Delta + \omega)^2 + (\kappa_m/2)^2} + \frac{\Delta - \omega}{(\Delta - \omega)^2 + (\kappa_m/2)^2} \right)
\end{eqnarray*}
and cause an additional damping rate of
\begin{equation*}
    \delta \kappa_{r1} = n_m g_{XZ1}^2 \frac{\omega_{r1}}{\omega} \left(\frac{\kappa_m}{(\Delta + \omega)^2 + (\kappa_m/2)^2} - \frac{\kappa_m}{(\Delta - \omega)^2 + (\kappa_m/2)^2} \right).
\end{equation*}
\end{widetext}
$\delta \omega_{r1}$ and $\delta \kappa_{r1}$ are analogous to the optomechanical spring effect and the optomechanical damping rate. The condition $\omega = \omega_{r1}$ can be assumed when $\sqrt{n_m} g_{XZ1} << \kappa$. 

\begin{figure}
\includegraphics[width = \columnwidth]{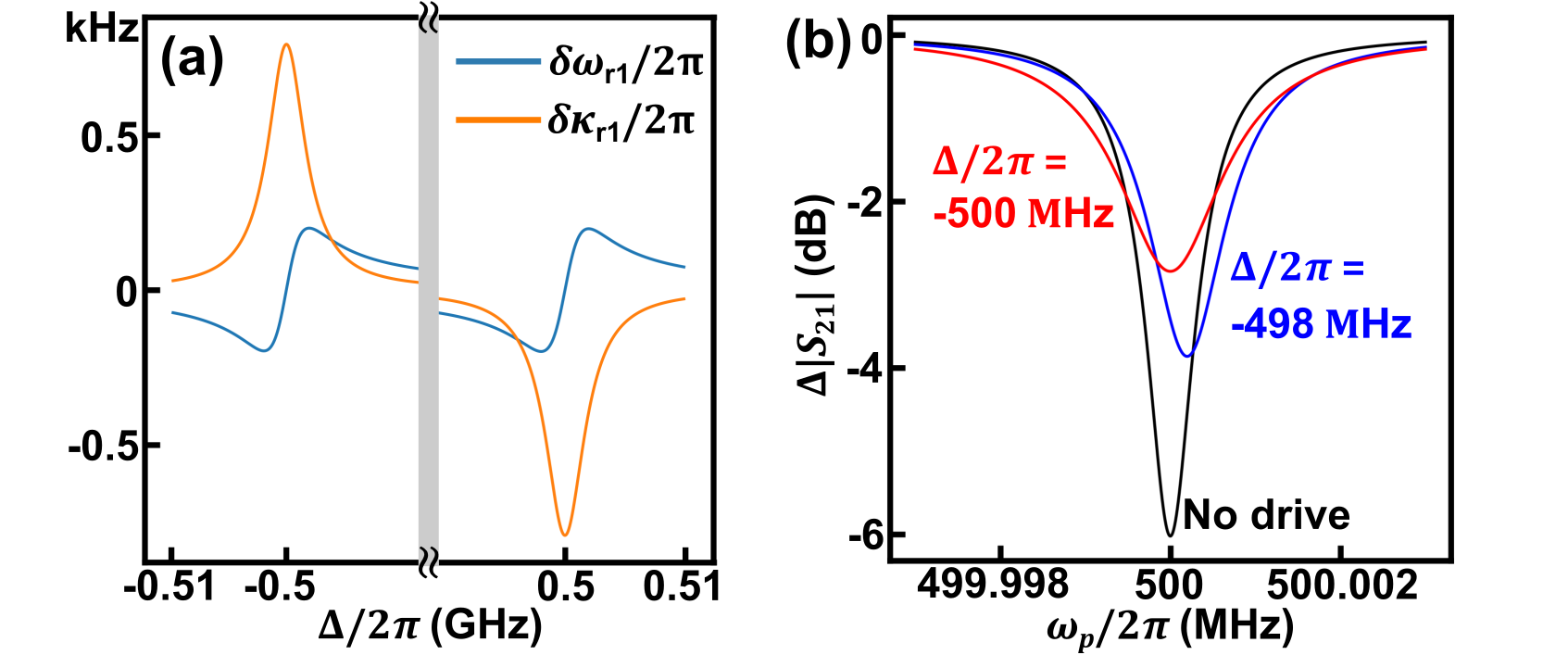}
\caption{(a) Change of fundamental mode's frequency $\delta \omega_{r1}$ and damping rate $\delta \kappa_{r1}$ caused by the backaction from the magnons vs the detuning of the MW drive from the Kittel mode frequency. The MW power in the magnon driving line is $-10$ dBm. When the detuning is $\pm 0.5$ GHz, the change of the fundamental mode damping rate is $\mp 792$ Hz. Assume Kittel mode frequency $\omega_m/2\pi = 20$ GHz, linewidth $\kappa_m/2\pi = 4$ MHz including the external loss rate to the CPW estimated to be $\kappa_{m,ext}/2\pi = 2$ MHz \cite{SIref}, fundamental resonator mode $\omega_{r1}/2\pi = 500$ MHz, $\kappa_{r1,i}/2\pi = \kappa_{r1,ext}/2\pi = 500$ Hz and $g_{XZ1}/2\pi = 12.835$ Hz. (b) Calculated transmission spectrum through the resonator feedline vs the probe frequency $\omega_p$ when there is no power in the magnon driving line (black); $P = -10$ dBm and $\Delta/2\pi = -500$ MHz (red); and $P = -10$ dBm and $\Delta/2\pi = -498$ MHz (blue).}
\label{top_CPW_dfdgamma_pic}
\end{figure}

Fig.~\ref{top_CPW_dfdgamma_pic}(a) shows $\delta \omega_{r1}$ and $\delta \kappa_{r1}$ as a function of $\Delta$ when the MW power in the feedline is $-10$ dBm. Here the magnon number $n_m$ follows the Lorentzian distribution as a function of $\Delta$ and its expression is given in SI. The effect of $\delta \omega_{r1}$ and $\delta \kappa_{r1}$ can be probed with the resonator feedline. 

Fig. \ref{top_CPW_dfdgamma_pic}(b) shows the resonator feedline transmission spectrum for different drivings in the magnon driving line. We can see that when the detuning is $\Delta = -\omega_{r1} = -500$ MHz, the magnons will create an additional internal damping of the resonator fundamental mode. Also, when $\Delta = -498$ MHz, the magnons will not only create an additional internal damping but also increase the resonance frequency of the fundamental mode.

We can create a mode squeezed state of the resonator fundamental by simultaneously driving the red and the blue magnon sidebands (two-tone drive) \cite{kronwald2013arbitrarily, wollman2015quantum, pirkkalainen2015squeezing, youssefi2023squeezed}. For a given red sideband driving power, we can adjust the blue sideband driving power such that the minimum variance of the squeezed quadrature $X_1$ is given by \cite{kronwald2013arbitrarily}:
\begin{align*}
    2\langle  \Delta\hat X_{1}^2 \rangle \approx \frac{\kappa_{r1}}{\kappa_m} (2n_{r1}^{th}+1) + \sqrt{\frac{2n_{r1}^{th}+1}{\cal C}}
\end{align*}
where the cooperativity $\cal C$ $= G_-^2 / (\kappa_{r1} \kappa_m) $, $G_- = \sqrt{n_-} g_{XZ1}$ and $n_-$ is the magnon number at the red sideband. Here we are assuming  $\cal C$ $>> 1$. 

Quantum squeezing can be achieved and detected with realistic parameters. For a 500~MHz resonator, the thermal photon occupation number is $\langle n^{th}_{r1} \rangle \approx 0.10$ at a dilution refrigerator base temperature of 10~mK. Using $\kappa_{r1}/2\pi = 1$ kHz and the red sideband MW power at the sample to be 0 dBm, we have $2\langle  \Delta\hat X_{1}^2 \rangle \approx 0.39$, corresponding to 4.1 dB squeezing. In this case the red detuned magnon number $n_m = 4.8 \times 10^7$, much smaller than the number of YIG spins $N = 3.8 \times 10^{11}$ \cite{SIref}. This squeezed state can be detected either with a homodyne circuit using the resonator feedline that is directly coupled to the resonator, with SC qubit \cite{dassonneville2021dissipative}, or with magnons through the magnon driving line with the method shown in \cite{kronwald2013arbitrarily, wollman2015quantum, pirkkalainen2015squeezing, youssefi2023squeezed} so that the resonator feedline can be eliminated to decrease $\kappa_{r1}$.

There are some other requirements for squeezing the fundamental mode. For example, the Kerr nonlinearity of magnons can be a problem for creating the resonator squeezed state. For example, if the frequency shift of magnons exceeds $\kappa_m$, it will complicate driving on resonance with the red (blue) sidebands. Kerr nonlinearity is represented by the term $K \hat{b}^{\dagger}\hat{b}\hat{b}^{\dagger}\hat{b}$ in the Hamiltonian, where $K \propto M_{eff}$ is the Kerr coefficient \cite{wang2016magnon}. In our top CPW geometry, we estimate that the Kerr coefficient from the shape anisotropy is $K = 0.01$~Hz. Therefore, when $\langle n_m \rangle = 4.8 \times 10^7$, the frequency shift of magnons will be 0.48~MHz, which is much smaller than $\kappa_m$. Although the Kerr nonlinearity will already be small, our scheme would work best if it is minimized. This can be satisfied by adjusting the strain on the YIG \cite{wang2024giant} to make $M_{eff} \approx 0$. Also, we need the inhomogeneous broadening of the magnon linewidth to be minimized, which can be done through homogeneous materials growth and elliptical pattering of YIG.

%\section{Alternative setup}

\begin{figure}
\includegraphics[width = \columnwidth]{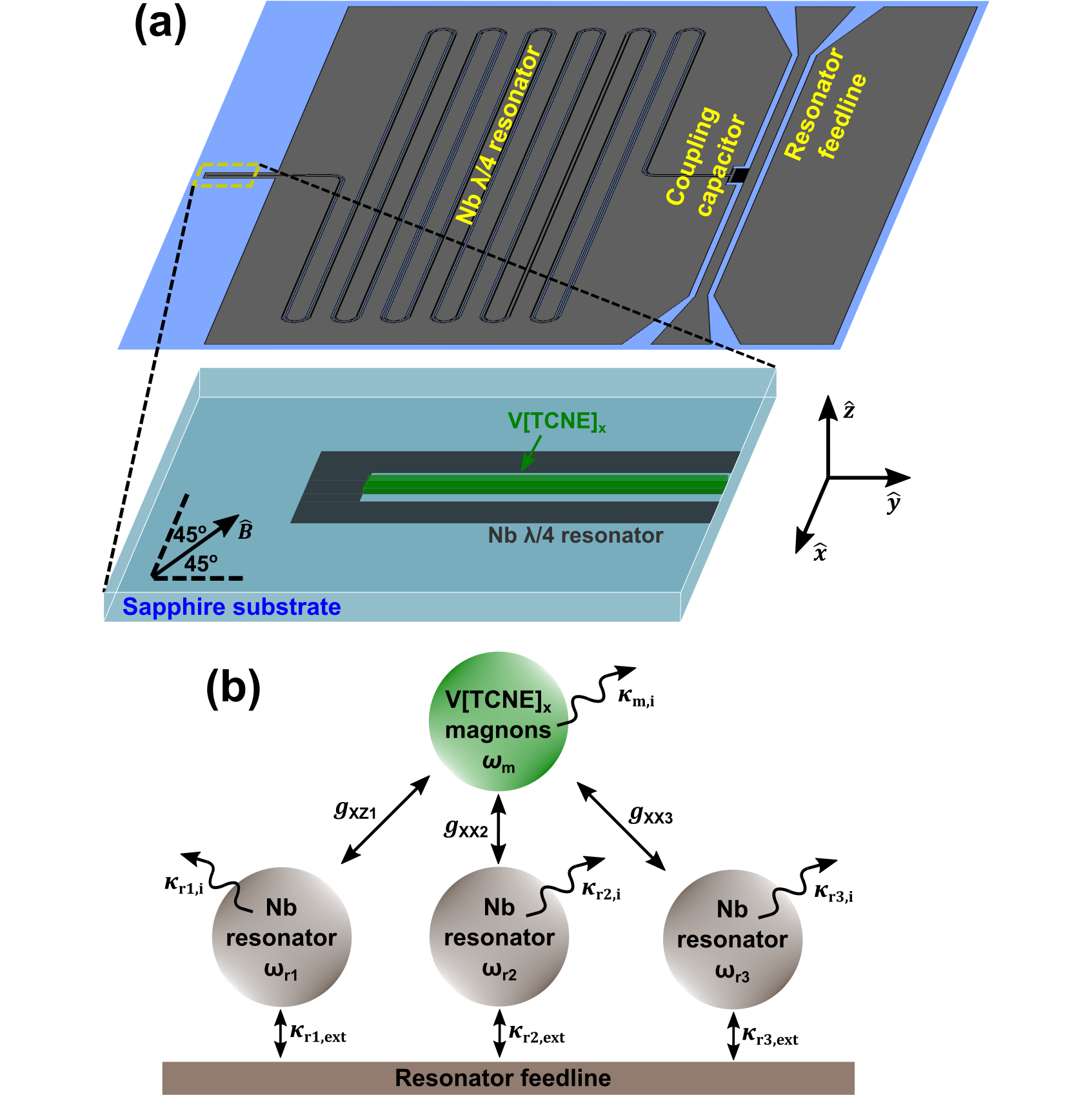}
\caption{(a) Schematic plot of the 45 degrees alignment setup and the zoomed in plot at the current antinode of the Nb $\lambda/4$ resonator. The geometry is similar to the top CPW setup except now the $\hat B$ is 45 degrees aligned to the resonator inductor wire and there's is no magnon driving line. Also, a bigger \VTNCE ~is used in this case as the magnon source. The static field magnitude $B$ is adjusted such that the Larmor frequency of \VTNCE ~magnons is around 20 GHz. With this geometry there's both XX and XZ coupling between magnons and the Nb resonator. Here we only consider the XZ interaction between magnons and the fundamental mode of the resonator and the XX coupling between magnons and the higher harmonic modes (19.5 GHz, and 20.5 GHz) of the resonator. The \VTNCE's dimension is 500 $\mu$m $ \times $ 5 $\mu$m $ \times$ 1 $\mu$m. (b) Scheme of all the dampings and couplings in the 45 degrees alignment setup.}
\label{45_geometry_pic}
\end{figure}

While the above experimental geometry is straightforward to analyze and conceptually simple, the crossed waveguide geometry will be more challenging to fabricate with low microwave loss and negligible cross-talk.  Therefore, we also propose an alternate geometry that is simpler to fabricate, would have decreased microwave driving power, and achieve the same (anti-)damping rate of the fundamental mode.  This geometry is shown in
Fig. \ref{45_geometry_pic}(a). Its is similar to the top CPW geometry except that this hybrid system is driven by the resonator feedline itself instead of the magnon driving line, and here we choose a larger volume of a different low-damping magnetic material -- \VTNCE -- to host the magnetic excitation. The \VTNCE ~is patterned directly on the central wire of the resonator as we have demonstrated previously~\cite{xu2024strong}, however, here the long axis is tilted 45 degrees with respect to the external static field.  This leads to both XX and XZ coupling between the \VTNCE~magnon mode and the resonator. In this configuration we can use the fundamental mode ($\omega_{r1}/2\pi = 500$ MHz) as the target for squeezing and two higher order harmonic modes ($\omega_{r2}/2\pi = 19.5$ GHz and $\omega_{r3}/2\pi = 20.5$ GHz) of the $\lambda/4$ resonator for driving. The external static field magnitude $B$ is adjusted such that the magnon frequency is around 20 GHz. 

The Hamiltonian of this system is:
\begin{equation*}
\begin{split}
{\cal H}/\hbar = &\omega_{r1} \hat{a}^\dagger_{r1} \hat{a}_{r1} + \omega_{r2} \hat{a}^\dagger_{r2} \hat{a}_{r2} + \omega_{r3} \hat{a}^\dagger_{r3} \hat{a}_{r3} 
\\&+\omega_{m}(B)\hat{b}^{\dagger}\hat{b} + g_{XX2}\left(\hat{b}^{\dagger}\hat{a}_{r2}+\hat{b}\hat{a}^\dagger_{r2}\right) 
\\&+g_{XX3}\left(\hat{b}^{\dagger}\hat{a}_{r3}+\hat{b}\hat{a}^\dagger_{r3}\right)- g_{XZ1}\hat{b}^{\dagger}\hat{b}\left(\hat{a}_{r1}+\hat{a}^\dagger_{r1}\right)
\label{Hamil_2}
\end{split}
\end{equation*}
where $g_{XZ1}$ is the XZ coupling strength between the resonator fundamental mode and the Kittel mode, and $g_{XX2}$ ($g_{XX3}$) is the XX coupling strength between the 19.5 GHz (20.5 GHz) resonator mode and the Kittel mode. The XX coupling between magnons and any other resonator mode are ignored because their detuning is much larger than their XX coupling strength. Also, the XZ coupling between magnons and any non-fundamental resonator mode is ignored when $|\Delta|/2\pi$ is much less than 1.5 GHz.

Fig. \ref{45_geometry_pic}(b) describes the interactions in this geometry. Note that the coupling capacitor between the CPW and the $\lambda/4$ resonator can be regarded as a high pass filter. Given the external damping rate of the higher harmonic mode $\kappa_{r2, ext}/2\pi = 2$ MHz, we estimate a much lower external damping rate of the fundamental mode $\kappa_{r1, ext}/2\pi = 1.25$ kHz \cite{SIref}.
Again, we assume a realistic value of the internal damping rate of the resonator fundamental mode $\kappa_{r1,i}/2\pi = 0.5$ kHz ($Q_i = 10^6$) so that $\kappa_{r1}/2\pi = 1.75$ kHz.

There are several advantages of this setup. Firstly, the magnon sidebands can be driven much more efficiently with cavity modes than with a broadband waveguide. While the top CPW geometry will have a too small efficiency to drive \VTNCE ~(which can be patterned more easily than YIG \cite{harberts2015chemical, zhu2016low, de2000cvd, pokhodnya2000thin, Fronig2015,Cheung2021}), in this geometry we can use either YIG or \VTNCE ~to do the sideband cooling without using too much MW power in the feedline that could potentially increase the mixing chamber temperature through the attenuators. Secondly, the bigger volume of the magnet means a smaller Kerr non-linearity. In this section, we focus on \VTNCE ~as the magnetic material.
\begin{comment}
    
In the rotating frame of the driving frequency, the equations of motion are
\begin{align*}
    \dot \alpha_{r2} &= -\frac{\kappa_{r2}}{2}\alpha_{r2} - i g_{XX2} \beta_m - i \Delta_{r2} \alpha_{r2}
    \\
    \dot \alpha_{r3} &= -\frac{\kappa_{r3}}{2}\alpha_{r3} - i g_{XX3} \beta_m - i \Delta_{r3} \alpha_{r3}
    \\
    \dot \beta_m &= -\frac{\kappa_m}{2} \beta_m + i(-\Delta_m + G\Phi) \beta_m - i g_{XX2} \alpha_{r2} - i g_{XX3} \alpha_{r3}
    \\
    C_1 \ddot \Phi &= -C_1 \omega_{r1}^2 \Phi - C_1 \kappa_{r1} \dot \Phi + \hbar G \left| \beta_m \right|^2
\end{align*}
where $\alpha_{ri} = \langle \hat a_{ri} \rangle$, and $\Delta_{r2} = \omega_{r2} - \omega_d$, $\Delta_{r3} = \omega_{r3} - \omega_d$ and $\Delta_m = \omega_m - \omega_d$ are the detunings from the driving frequency. Following the same procedure in section \ref{prop_setup}, we can get $\delta \omega_{r1}$ and $\delta \kappa_{r1}$ as a function of external field $B$ and driving frequency $\omega_d$ (SI).
\end{comment}

\begin{figure}
\includegraphics[width = \columnwidth]{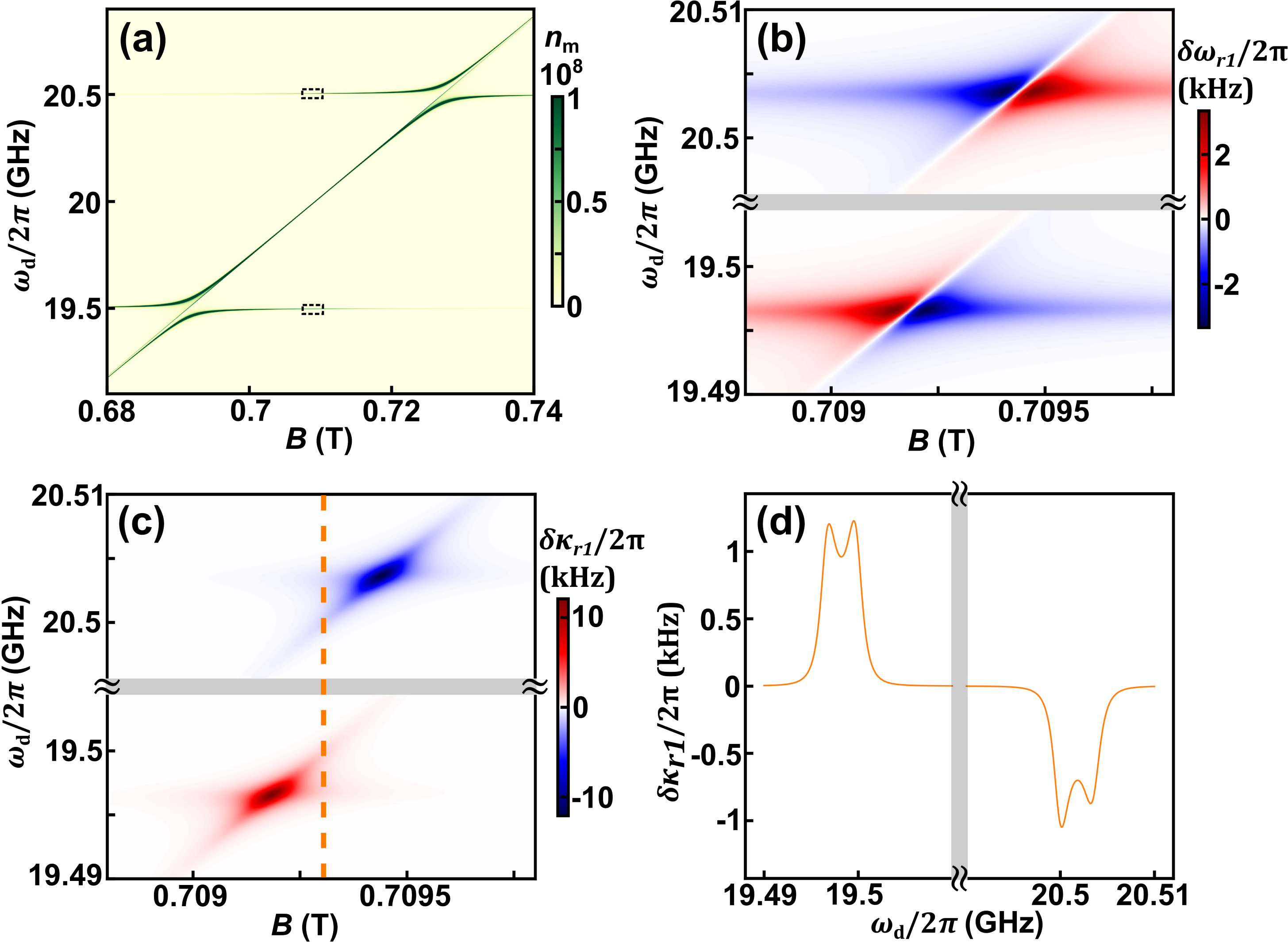}
\caption{(a) Field-frequency plot of magnon number when the MW power is $-30$ dBm in case of \VTNCE ~in the 45 degrees alignment geometry (Spots where $n_m \ge 10^{8}$ are represented by the same color). The dashed black boxes are the spots where the backaction damping/antidamping rate has the biggest amplitude. (b) The backaction frequency shift in the dashed black boxes region. (c) The backaction damping rate in the dashed black boxes region. In this case the cooling efficiency is much higher in comparison with Fig. \ref{top_CPW_dfdgamma_pic} (especially when $\kappa_{m}<<\omega_{r1}$). (d) Linecut at 0.709305 T in (c) (orange dashed line), where $\omega_m/2\pi = 20$ GHz. Parameter used: $\omega_{r1}/2\pi = 500$ MHz, $\omega_{r2}/2\pi = 19.5$ GHz, $\omega_{r3}/2\pi = 20.5$ GHz, $\kappa_{r2,i}/2\pi = \kappa_{r3,i}/2\pi = 20$ kHz, $\kappa_{r2, ext}/2\pi = 2$ MHz, $\kappa_{r3, ext}/2\pi = 2.21$ MHz, $\kappa_m/2\pi = \kappa_{m,i}/2\pi = 2$ MHz, $g_{XZ1}/2\pi = 9.076$ Hz, $g_{XX2}/2\pi = 41.5$ MHz, $g_{XX3}/2\pi = 42.6$ MHz (see the SI for the derivation of these values \cite{SIref}).}
\label{YIG_VTCNE_frequency_damping_pic}
\end{figure}

Following an analytical procedure similar to the one above, we calculate properties of the system including $\delta \omega_{r1}$ and $\delta \kappa_{r1}$ as a function of external field $B$ and driving frequency $\omega_d$ \cite{SIref}.
Fig.~\ref{YIG_VTCNE_frequency_damping_pic} shows the field-frequency plot of magnon number, the backaction frequency shift and the backaction damping rate when the MW power is $-30$ dBm in case of \VTNCE ~in the 45 degrees alignment geometry. In these plots we can see that the magnitude of the backaction damping rate reaches a maximum when the sidebands of the middle branch (middle eigenfrequency) cross the upper or lower branches of the magnon number plot. Comparing to Fig. \ref{top_CPW_dfdgamma_pic}, we can reach a higher (anti-)damping rate and thus a higher squeezing rate with the same microwave power at the sample. Fig. \ref{YIG_VTCNE_frequency_damping_pic}(d) shows the backaction damping rate as a function of driving frequency when \VTNCE's Larmor frequency is 20 GHz, in which we can create the squeezed state by driving at $\Delta/2\pi = \pm500$ MHz.

There are a few caveats for this geometry.  The higher harmonic modes exist in the same $\lambda/4$ resonator as the 500 MHz fundamental mode. In this case, the excitation in the higher harmonic modes modes may cause decoherence of the fundamental resonator mode and the heat dissipation can increase $n_{r1}^{th}$. If this proves to be a problem, one can use the geometry of Fig. \ref{top_CPW_geometry_pic} except change the magnon driving line into a resonator with two modes for sideband driving (19.5 GHz and 20.5 GHz).

%In our case of $\kappa_{r2} << \omega_{r1}$, it is better to use the 45 degree geometry to drive the magnons more efficiently and achieve a higher backaction (anti)damping rate with the same MW power. Such cooling and squeezing with a coupled system can also be achieved with other microwave optomechanical setups at low T \cite{wollman2015quantum, pirkkalainen2015squeezing, youssefi2023squeezed} \icite to decrease the microwave power in the feedline. 

%\section{conclusion}

We propose two geometries to realize the beam splitter interaction (XZ coupling) in cavity magnonics. The 45 degree alignment setup uses both XX and XZ coupling between magnons and MW photons to drive the magnon sidebands more efficiently, thus reducing the MW power at the sample. This advantage is obvious especially when $\kappa_{m}<<\omega_{r1}$. Our analytical results show that we can not only make the backaction damping (anti-damping) rate larger than the bare MW resonator damping rate, but we can also achieve quantum squeezing of the MW resonator when the magnon number is much smaller than the number of spins. Our proposal provides another fundamental building block for manipulating hybrid quantum systems.

\begin{acknowledgments}

This work was supported the Department of Energy Office of Science, Basic Energy Sciences Quantum Information Sciences program (DE-SC0019250).

\end{acknowledgments}

\nocite{*}

\bibliography{OM_BIB}% Produces the bibliography via BibTeX.

\newpage

\clearpage

%\title{Supplementary Information for Cooling and Squeezing a Microwave Cavity State with Magnons Using a Beam Splitter Interaction}

%\author{Qin Xu}
%\affiliation{Department of Physics, Cornell University, Ithaca NY 14853}
%\author{Gregory D. Fuchs}
%\affiliation{School of Applied and Engineering Physics, Cornell University, Ithaca NY 14853}


\begin{widetext}

%\maketitle
\begin{center}
    \begin{large}
        \textbf{Supplementary Information for Cooling and Squeezing a Microwave Cavity State with Magnons Using a Beam Splitter Interaction}
    \end{large}
\end{center}

\section{1. Parameters used}

Parameters assumed in the main text are listed in Table \ref{table top CPW} (for top CPW setup) and \ref{taqble 45} (for 45 degrees alignment setup) respectively. Some of them are calculated in sections below. 

\begin{table}[h]
\begin{tabular}{cccccc}
\hline
\multicolumn{1}{|c|}{Parameters}  & \multicolumn{1}{c|}{Values} & \multicolumn{4}{c|}{Meaning} \\ \hline
\multicolumn{1}{|c|}{$\omega_{r1}/2\pi$}  & \multicolumn{1}{c|}{$500$ MHz} & \multicolumn{4}{c|}{Oscillation frequency of the $\lambda/4$ resonator fundamental mode} \\ \hline
\multicolumn{1}{|c|}{$\omega_{m}/2\pi$}  & \multicolumn{1}{c|}{$20$ GHz} & \multicolumn{4}{c|}{Oscillation frequency of magnons} \\ \hline
\multicolumn{1}{|c|}{$\kappa_{r1,i}/2\pi$}  & \multicolumn{1}{c|}{$500$ Hz} & \multicolumn{4}{c|}{Internal decay rate of the $\lambda/4$ resonator fundamental mode} \\ \hline
\multicolumn{1}{|c|}{$\kappa_{m,i}/2\pi$}  & \multicolumn{1}{c|}{$2$ MHz} & \multicolumn{4}{c|}{Internal decay rate of magnons} \\ \hline
\multicolumn{1}{|c|}{$\kappa_{r1,ext}/2\pi$}  & \multicolumn{1}{c|}{$500$ Hz} & \multicolumn{4}{c|}{External decay rate of the $\lambda/4$ resonator fundamental mode} \\ \hline
\multicolumn{1}{|c|}{$\kappa_{m,ext}/2\pi$}  & \multicolumn{1}{c|}{$2$ MHz} & \multicolumn{4}{c|}{External decay rate of magnons} \\ \hline
\multicolumn{1}{|c|}{$g_{XZ1}/2\pi$}  & \multicolumn{1}{c|}{$12.835$ Hz} & \multicolumn{4}{c|}{\makecell{XZ coupling rate between the $\lambda/4$ resonator fundamental mode \\and magnons}} \\ \hline
\end{tabular}
\caption{Parameters used in the top CPW setup}
    \label{table top CPW}
\end{table}

\begin{table}[]
\begin{tabular}{cccccc}
\hline
\multicolumn{1}{|c|}{Parameters}  & \multicolumn{1}{c|}{Values} & \multicolumn{4}{c|}{Meaning} \\ \hline
\multicolumn{1}{|c|}{$\omega_{r1}/2\pi$}  & \multicolumn{1}{c|}{$500$ MHz} & \multicolumn{4}{c|}{Oscillation frequency of the $\lambda/4$ resonator fundamental mode} \\ \hline
\multicolumn{1}{|c|}{$\omega_{r2}/2\pi$}  & \multicolumn{1}{c|}{$19.5$ GHz} & \multicolumn{4}{c|}{Oscillation frequency of one of the $\lambda/4$ resonator higher order mode} \\ \hline
\multicolumn{1}{|c|}{$\omega_{r3}/2\pi$}  & \multicolumn{1}{c|}{$20.5$ GHz} & \multicolumn{4}{c|}{Oscillation frequency of another $\lambda/4$ resonator higher order mode} \\ \hline
\multicolumn{1}{|c|}{$\kappa_{r1,i}/2\pi$}  & \multicolumn{1}{c|}{$500$ Hz} & \multicolumn{4}{c|}{Internal decay rate of the $\lambda/4$ resonator fundamental mode} \\ \hline
\multicolumn{1}{|c|}{$\kappa_{r2,i}/2\pi$}  & \multicolumn{1}{c|}{$20$ kHz} & \multicolumn{4}{c|}{Internal decay rate of the $\lambda/4$ resonator $19.5$ GHz mode} \\ \hline
\multicolumn{1}{|c|}{$\kappa_{r3,i}/2\pi$}  & \multicolumn{1}{c|}{$20$ kHz} & \multicolumn{4}{c|}{Internal decay rate of the $\lambda/4$ resonator $20.5$ GHz mode} \\ \hline
\multicolumn{1}{|c|}{$\kappa_{m,i}/2\pi$}  & \multicolumn{1}{c|}{$2$ MHz} & \multicolumn{4}{c|}{Internal decay rate of magnons} \\ \hline
\multicolumn{1}{|c|}{$\kappa_{r1,ext}/2\pi$}  & \multicolumn{1}{c|}{$1.25$ kHz} & \multicolumn{4}{c|}{External decay rate of the $\lambda/4$ resonator fundamental mode} \\ \hline
\multicolumn{1}{|c|}{$\kappa_{r2,ext}/2\pi$}  & \multicolumn{1}{c|}{$2$ MHz} & \multicolumn{4}{c|}{External decay rate of the $\lambda/4$ resonator $19.5$ GHz mode} \\ \hline
\multicolumn{1}{|c|}{$\kappa_{r3,ext}/2\pi$}  & \multicolumn{1}{c|}{$2.21$ MHz} & \multicolumn{4}{c|}{External decay rate of the $\lambda/4$ resonator $20.5$ GHz mode} \\ \hline
\multicolumn{1}{|c|}{$g_{XZ1}/2\pi$}  & \multicolumn{1}{c|}{$9.076$ Hz} & \multicolumn{4}{c|}{\makecell{XZ coupling rate between the $\lambda/4$ resonator fundamental mode \\and magnons}} \\ \hline
\multicolumn{1}{|c|}{$g_{XX2}/2\pi$}  & \multicolumn{1}{c|}{$41.5$ MHz} & \multicolumn{4}{c|}{\makecell{XX coupling rate between the $\lambda/4$ resonator $19.5$ GHz mode \\and magnons}} \\ \hline
\multicolumn{1}{|c|}{$g_{XX3}/2\pi$}  & \multicolumn{1}{c|}{$42.6$ MHz} & \multicolumn{4}{c|}{\makecell{XX coupling rate between the $\lambda/4$ resonator $20.5$ GHz mode \\and magnons}} \\ \hline
\end{tabular}
\caption{Parameters used in the 45 degrees alignment setup}
    \label{taqble 45}
\end{table}

\section{2. Equations of motion of the top CPW setup}

As shown in the main text, the Hamiltonian of this system is
\begin{equation*}
    {\cal H}/\hbar = \omega_{r1} \hat{a}^\dagger_{r1} \hat{a}_{r1}+\omega_{m}(B)\hat{b}^{\dagger}\hat{b} - g_{XZ1}\hat{b}^{\dagger}\hat{b}\left(\hat{a}_{r1}+\hat{a}^\dagger_{r1}\right),
    \label{Hamil_1}
\end{equation*}
where the first 2 terms are the energies of the superconducting resonator's fundamental mode (denoted by subscript $r1$) and the Kittel mode of the magnet (subscript $m$), respectively. The third term is the XZ coupling between the resonator's fundamental mode and the Kittel mode with coupling strength $g_{XZ1} = G\Phi_{ZPF}$, where $\Phi_{ZPF} \propto I_{ZPF}$ is the zero point fluctuation of fundamental mode flux and $G$ is the magnon frequency shift per unit flux.

The XZ coupling between the two systems means a change in one of the quadratures (such as the position or momentum in the case of a mechanical oscillator) of the first system will change the frequency of the second. In our system, $g_{XZ1}$ is equal to the Kittel mode frequency shift caused by the zero point fluctuation of the resonator's flux, $\Phi_{ZPF}$
\begin{equation*}
\begin{split}
    g_{XZ1} &= \frac{\partial \omega_{m}}{\partial \Phi} \Phi_{ZPF}
    = \frac{\partial \omega_{m}}{\partial I} I_{ZPF} = \frac{\partial \omega_{m}}{\partial I} \sqrt{\frac{2\hbar}{\pi Z_0}}\omega_{r1} \\
    &= \gamma \frac{\mu_0}{2w} \sqrt{\frac{2\hbar}{\pi Z_0}}\omega_{r1}
    = 2\pi \times 12.835\, \text{Hz},
\end{split}
\end{equation*}
where we use $I$ to indicate the current at the current antinode of the $\lambda/4$ resonator and $I_{ZPF} = \sqrt{\frac{2\hbar}{\pi Z_0}}\omega_{r1}$ (see section 5).

In the classical regime, we can denote the magnon amplitude $\beta_m[t] = \langle \hat b[t] \rangle$, the resonator amplitude $\alpha_{r1}[t] = \langle \hat a_{r1}[t] \rangle$. Also, we denote the flux of the resonator as $\Phi[t] = \langle \hat \Phi[t] \rangle = \Phi_{ZPF}(\langle\hat a_{r1}[t] + \hat a_{r1}^{\dagger}[t]\rangle)$. We define $L_1 = \sqrt{\frac{Z_0}{\omega_{r1}}}$ and $C_1 = \sqrt{\frac{1}{\omega_{r1} Z_0}}$. Similar to an optomechanical system \cite{aspelmeyer2014cavity}, the equations of motion for an XZ coupled system is:
\begin{equation}
\begin{split}
    \dot \beta_m &= -\frac{\kappa}{2} \beta_m + i(\Delta + G\Phi) \beta_m 
    \\
    C_1 \ddot \Phi &= -C_1 \omega_{r1}^2 \Phi - C_1 \kappa_{r1} \dot \Phi + \hbar G \left| \beta_m \right|^2
    \label{classical_time}
\end{split}
\end{equation}
where we have used the rotating frame of the external drive frequency $\omega_d$ and $\Delta = \omega_d - \omega_m$. 

In a linearized treatment we can write the magnon amplitude $\beta_m[t] = \bar\beta_m + \delta \beta_m[t]$, where $\bar\beta_m$ is the magnon amplitude averaged over time \cite{aspelmeyer2014cavity}. The equations of motion then becomes
\begin{equation}
\begin{split}
    \delta \dot \beta_m &= (i \Delta -\frac{\kappa}{2}) \delta \beta_m + i G \bar \beta_m \Phi
    \\
    C_1 \ddot \Phi &= -C_1 \omega_{r1}^2 \Phi - C_1 \kappa_{r1} \dot \Phi + \hbar G (\bar \beta_m^* \delta \beta_m + \bar \beta_m \delta \beta_m^*).
    \label{classical_time_linear}
\end{split}
\end{equation}

Note that from equations \ref{classical_time} to \ref{classical_time_linear}, we have ignored the constant terms on the right hand side ($\kappa \bar \beta_m$, $\Delta \bar \beta_m$, and $\bar\beta_m \bar\beta_m^*$) because these will only cause a constant displacement of $\Phi[t]$. Also, we ignore $\delta \beta_m[t]\delta \beta_m^*[t]$ and $\Phi[t]\delta \beta_m[t]$ under the rotating wave approximation. To transform into the frequency domain, we assume oscillatory solutions $\delta\beta_m = \delta\beta_m[\omega] e^{-i\omega t}$ and $\Phi = \Phi[\omega] e^{-i\omega t}$. In that case we have
\begin{align*}
    - i \omega \delta \beta_m[\omega] =& (i \Delta -\frac{\kappa}{2}) \delta \beta_m[\omega] + i G \bar \beta_m \Phi[\omega]
    \\
    - C_1 \omega^2 \Phi[\omega] =& -C_1 \omega_{r1}^2 \Phi[\omega] + i \omega C_1 \kappa_{r1} \Phi[\omega] + \hbar G (\bar \beta_m^* \delta \beta_m[\omega] 
    \\ &+ \bar \beta_m \delta \beta_m^*[\omega]).
\end{align*}
So
\begin{align*}
    \delta \beta_m [\omega] &= \frac{iG \bar \beta_m}{\kappa/2 - i(\Delta + \omega)} \Phi [\omega],
\end{align*}
\begin{align*}
    \delta \beta_m^* [\omega] &= \delta \beta_m [-\omega]^* = \frac{-iG \bar \beta_m}{\kappa/2 + i(\Delta - \omega)} \Phi [\omega],
\end{align*}
and
\begin{align*}
    \hbar G (\bar \beta_m^* \delta \beta_m[\omega] + \bar \beta_m \delta \beta_m^*[\omega])=&\hbar G \left(\frac{iG |\bar \beta_m|^2}{\kappa/2 - i(\Delta + \omega)} - \frac{iG |\bar \beta_m|^2}{\kappa/2 + i(\Delta - \omega)} \right) \Phi [\omega]
    \\ = &-2 C_1 \omega_{r1} n g_{XZ1}^2 \left(\frac{1}{(\Delta + \omega) + i \kappa/2} + \frac{1}{(\Delta - \omega) - i \kappa/2} \right) \Phi [\omega].
\end{align*}

With the magnetic material inductively coupled at the current antinode, the fundamental resonator mode inductance $L_{eff}$ will be given by
\begin{align*}
    \frac{1}{L_{eff}} &= C_1 \omega_{r1}^2 + 2 C_1 \omega_{r1} n g_{XZ1}^2 \left(\frac{1}{(\Delta + \omega) + i \kappa/2} + \frac{1}{(\Delta - \omega) - i \kappa/2} \right)
    \\
    &= \frac{1}{L_1} + 2 C_1 \omega_{r1} n g_{XZ1}^2 \left(\frac{1}{(\Delta + \omega) + i \kappa/2} + \frac{1}{(\Delta - \omega) - i \kappa/2} \right).
\end{align*}
So the backaction will shift the resonator frequency by
\begin{equation*}
    \delta \omega_{r1} = n_m g_{XZ1}^2 \frac{\omega_{r1}}{\omega} \left(\frac{\Delta + \omega}{(\Delta + \omega)^2 + (\kappa/2)^2} + \frac{\Delta - \omega}{(\Delta - \omega)^2 + (\kappa/2)^2} \right)
\end{equation*}
and cause an additional damping rate of
\begin{equation*}
    \delta \kappa_{r1} = n_m g_{XZ1}^2 \frac{\omega_{r1}}{\omega} \left(\frac{\kappa}{(\Delta + \omega)^2 + (\kappa/2)^2} - \frac{\kappa}{(\Delta - \omega)^2 + (\kappa/2)^2} \right).
\end{equation*}
$\delta \omega_{r1}$ and $\delta \kappa_{r1}$ are analagous to the optomechanical spring effect and the optomechanical damping rate. We can assume $\omega = \omega_{r1}$ when $\sqrt{n_m} g_{XZ1} << \kappa$.

\section{3. Equations of motion of the 45 degrees alignment setup}
\label{cool_45}

We use the fundamental mode ($\sim$ 500 MHz, denoted by $r_1$) and two of the higher order harmonic modes ($\omega_{r2}/2\pi = 19.5$ GHz and $\omega_{r3}/2\pi = 20.5$ GHz) of the $\lambda/4$ resonator. The external static field magnitude $B$ is adjusted such that the magnon frequency is around 20~GHz. 

The Hamiltonian of this system is
\begin{align}
{\cal H}/\hbar = &\omega_{r1} \hat{a}^\dagger_{r1} \hat{a}_{r1} + \omega_{r2} \hat{a}^\dagger_{r2} \hat{a}_{r2} + \omega_{r3} \hat{a}^\dagger_{r3} \hat{a}_{r3} 
+\omega_{m}(B)\hat{b}^{\dagger}\hat{b} + g_{XX2}\left(\hat{b}^{\dagger}\hat{a}_{r2}+\hat{b}\hat{a}^\dagger_{r2}\right) 
\\&+g_{XX3}\left(\hat{b}^{\dagger}\hat{a}_{r3}+\hat{b}\hat{a}^\dagger_{r3}\right)- g_{XZ1}\hat{b}^{\dagger}\hat{b}\left(\hat{a}_{r1}+\hat{a}^\dagger_{r1}\right),
\label{Hamil_2}
\end{align}
where $g_{XZ1}$ is the XZ coupling strength between the resonator fundamental mode and the Kittel mode, and $g_{XX2}$ ($g_{XX3}$) is the XX coupling strength between the 19.5~GHz (20.5~GHz) resonator mode and the Kittel mode.
In the rotating frame of the driving frequency, the equations of motion are
\begin{align*}
    \dot \alpha_{r2} &= -\frac{\kappa_{r2}}{2}\alpha_{r2} - i g_{XX2} \beta_m - i \Delta_{r2} \alpha_{r2}
    \\
    \dot \alpha_{r3} &= -\frac{\kappa_{r3}}{2}\alpha_{r3} - i g_{XX3} \beta_m - i \Delta_{r3} \alpha_{r3}
    \\
    \dot \beta_m &= -\frac{\kappa_m}{2} \beta_m + i(-\Delta_m + G\Phi) \beta_m - i g_{XX2} \alpha_{r2} - i g_{XX3} \alpha_{r3}
    \\
    C_1 \ddot \Phi &= -C_1 \omega_{r1}^2 \Phi - C_1 \kappa_{r1} \dot \Phi + \hbar G \left| \beta_m \right|^2,
\end{align*}
where $\alpha_{ri} = \langle \hat a_{ri} \rangle$, and $\Delta_{r2} = \omega_{r2} - \omega_d$, $\Delta_{r3} = \omega_{r3} - \omega_d$ and $\Delta_m = \omega_m - \omega_d$ are the detunings from the driving frequency.

In a linearized treatment we can write the 19.5 GHz resonator mode amplitude as $\alpha_{r2}[t] = \bar\alpha_{r2} + \delta \alpha_{r2}[t]$, the 20.5 GHz resonator mode amplitude as $\alpha_{r3}[t] = \bar\alpha_{r3} + \delta \alpha_{r3}[t]$ and magnon amplitude as $\beta_m[t] = \bar\beta_m + \delta \beta_m[t]$.  We then find
\begin{align*}
    \delta \dot \alpha_{r2} &= -\frac{\kappa_{r2}}{2} \delta \alpha_{r2} - i g_{XX2} \delta \beta_m - i \Delta_{r2} \delta \alpha_{r2}
    \\
    \delta \dot \alpha_{r3} &= -\frac{\kappa_{r3}}{2} \delta \alpha_{r3} - i g_{XX3} \delta \beta_m - i \Delta_{r3} \delta \alpha_{r3}
    \\
    \delta \dot \beta_m &= -\frac{\kappa_m}{2} \delta \beta_m - i \Delta_m \delta \beta_m + i G\Phi \bar\beta_m - i g_{XX2} \delta \alpha_{r2} - i g_{XX3} \delta \alpha_{r3}
    \\
    C_1 \ddot \Phi &= -C_1 \omega_{r1}^2 \Phi - C_1 \kappa_{r1} \dot \Phi + \hbar G (\bar \beta_m^* \delta \beta_m + \bar \beta_m \delta \beta_m^*).
\end{align*}

In the frequency domain we have:
\begin{align*}
    - i \omega \delta \alpha_{r2}[\omega]&= -\frac{\kappa_{r2}}{2} \delta \alpha_{r2}[\omega]- i g_{XX2} \delta \beta_m[\omega] - i \Delta_{r2} \delta \alpha_{r2}[\omega]
    \\
    - i \omega \delta \alpha_{r3}[\omega]&= -\frac{\kappa_{r3}}{2} \delta \alpha_{r3}[\omega]- i g_{XX3} \delta \beta_m[\omega] - i \Delta_{r3} \delta \alpha_{r3}[\omega]
    \\
    - i \omega \delta \beta_m[\omega] &= -\frac{\kappa_m}{2} \delta \beta_m[\omega] - i \Delta_m \delta \beta_m[\omega] + i G\Phi[\omega] \bar\beta_m - i g_{XX2} \delta \alpha_{r2}[\omega]- i g_{XX3} \delta \alpha_{r3}[\omega]
    \\
    - C_1 \omega^2 \Phi[\omega] &= -C_1 \omega_{r1}^2 \Phi[\omega] + i \omega C_1 \kappa_{r1} \Phi[\omega] + \hbar G (\bar \beta_m^* \delta \beta_m[\omega] + \bar \beta_m \delta \beta_m^*[\omega]).
\end{align*}
So:
\begin{align*}
    - i g_{XX2} \delta \alpha_{r2}[\omega]&= \frac{g_{XX2}}{i (\omega - \Delta_{r2}) - \kappa_{r2}/2} \delta \beta_m[\omega]
    \\
     - i g_{XX3} \delta \alpha_{r3}[\omega]&= \frac{g_{XX3}}{i (\omega - \Delta_{r3}) - \kappa_{r3}/2} \delta \beta_m[\omega]
    \\
    \delta \beta_m [\omega] &= \frac{iG \bar \beta_m}{\kappa_m/2 - i(\omega - \Delta_m) - \frac{g_{XX2}^2}{i (\omega - \Delta_{r2}) - \kappa_{r2}/2} - \frac{g_{XX3}^2}{i (\omega - \Delta_{r3}) - \kappa_{r3}/2}} \Phi [\omega]
\end{align*}
and
\begin{align*}
    &\hbar G (\bar \beta_m^* \delta \beta_m[\omega] + \bar \beta_m \delta \beta_m^*[\omega]) 
    \\=&-2 C_1 \omega_{r1} n_m g_{XZ}^2 \frac{1}{(-\Delta_m + \omega) + i \kappa_m/2 - \frac{g_{XX2}^2}{(\omega - \Delta_{r2}) + i \kappa_{r2}/2}- \frac{g_{XX3}^2}{(\omega - \Delta_{r3}) + i\kappa_{r3}/2}} \Phi[\omega]
    \\&-2 C_1 \omega_{r1} n_m g_{XZ}^2 \frac{1}{(-\Delta_m - \omega) - i \kappa_m/2 - \frac{g_{XX2}^2}{(-\omega - \Delta_{r2}) - i \kappa_{r2}/2}- \frac{g_{XX3}^2}{(-\omega - \Delta_{r3}) - i \kappa_{r3}/2}} \Phi[\omega].
\end{align*}

With the magnetic material at the current antinode, the fundamental mode inductance $L_{eff}$ will be given by:
\begin{align*}
    \frac{1}{L_{eff}} = \frac{1}{L_1} + 2 C_1 \omega_{r1} n g_{XZ}^2 \left(\frac{1}{A + i C} + \frac{1}{B - i D} \right)
\end{align*}
where
\begin{align*}
    A &= -\Delta_m + \omega - \frac{g_{XX2}^2 (-\Delta_{r2} + \omega)}{\kappa_{r2}^2/4 + (-\Delta_{r2} + \omega)^2} - \frac{g_{XX3}^2 (-\Delta_{r3} + \omega)}{\kappa_{r3}^2/4 + (-\Delta_{r3} + \omega)^2}
    \\
    B &= -\Delta_m - \omega - \frac{g_{XX2}^2 (-\Delta_{r2} - \omega)}{\kappa_{r2}^2/4 + (-\Delta_{r2} - \omega)^2} - \frac{g_{XX3}^2 (-\Delta_{r3} - \omega)}{\kappa_{r3}^2/4 + (-\Delta_{r3} - \omega)^2}
    \\
    C &= \kappa_{m}/2 + \frac{g_{XX2}^2 \kappa_{r2}/2}{\kappa_{r2}^2/4 + (-\Delta_{r2} + \omega)^2} + \frac{g_{XX3}^2 \kappa_{r3}/2}{\kappa_{r3}^2/4 + (-\Delta_{r3} + \omega)^2}
    \\
    D &= \kappa_{m}/2 + \frac{g_{XX2}^2 \kappa_{r2}/2}{\kappa_{r2}^2/4 + (-\Delta_{r2} - \omega)^2} + \frac{g_{XX3}^2 \kappa_{r3}/2}{\kappa_{r3}^2/4 + (-\Delta_{r3} - \omega)^2}
\end{align*}
So the magnons will shift the CPW resonator frequency by
\begin{align*}
    \delta \omega_{r1} = n_m g_{XZ}^2 \frac{\omega_{r1}}{\omega} \left(\frac{A}{A^2 + C^2} + \frac{B}{B^2 + D^2} \right)
\end{align*}
and cause an additional damping rate of
\begin{align*}
    \delta \kappa_{r1} = n_m g_{XZ}^2 \frac{\omega_{r1}}{\omega} \left(\frac{2C}{A^2 + C^2} + \frac{2D}{B^2 + D^2} \right).
\end{align*}

Now we need the field-frequency dependence of magnon number when the driving frequency is around 20 GHz. From the  Langevin equations, when both the 19.5~GHz and 20.5~GHz resonator modes are coupled to the magnons and both are excited by the waveguide with power $P$ at frequency $\omega_d$, the number of photons stored in each of the MW resonator modes is
\begin{align*}
    \langle n_{r2} \rangle &= \left| \langle \alpha_{r2} \rangle \right|^2 = \frac{2PQ_{r2,c}}{\hbar \omega_d^2} \left| \frac{(\kappa_{r2,ext} / 2)}{i(\omega_d - \omega_{r2})-\kappa_{r2} / 2 + \frac{g_{XX2}^2}{i(\omega_d - \omega_{m})-\kappa_{m} / 2 + \frac{g_{XX3}^2}{i(\omega_d - \omega_{r3})-\kappa_{r3} / 2}}}\right|^2
    \\
    \langle n_{r3} \rangle &= \left| \langle \alpha_{r3} \rangle \right|^2 = \frac{2PQ_{r3,c}}{\hbar \omega^2_d} \left| \frac{(\kappa_{r3,ext} / 2)}{i(\omega_d - \omega_{r3})-\kappa_{r3} / 2 + \frac{g_{XX3}^2}{i(\omega_d - \omega_{m})-\kappa_{m} / 2 + \frac{g_{XX2}^2}{i(\omega_d - \omega_{r2})-\kappa_{r2} / 2}}}\right|^2,
\end{align*}
where the MW resonator oscillation amplitude is
\begin{align*}
    \langle \alpha_{r2} \rangle =& \sqrt{\frac{2PQ_{r2,c}}{\hbar \omega_d^2}} \frac{(\kappa_{r2,ext} / 2)}{i(\omega_d - \omega_{r2})-\kappa_{r2} / 2 + \frac{g_{XX2}^2}{i(\omega_d - \omega_{m})-\kappa_{m} / 2 + \frac{g_{XX3}^2}{i(\omega_d - \omega_{r3})-\kappa_{r3} / 2}}}
    \\
    \langle \alpha_{r3} \rangle =& \sqrt{\frac{2PQ_{r3,c}}{\hbar \omega_d^2}} \frac{(\kappa_{r3,ext} / 2)}{i(\omega_d - \omega_{r3})-\kappa_{r3} / 2 + \frac{g_{XX3}^2}{i(\omega_d - \omega_{m})-\kappa_{m} / 2 + \frac{g_{XX2}^2}{i(\omega_d - \omega_{r2})-\kappa_{r2} / 2}}}
\end{align*}
and $Q_{ri,c} = \omega_{ri} / \kappa_{ri,ext}$ are the external quality factors of the higher order harmonic modes of the resonator.
So the magnon amplitude is:
\begin{align*}
     \langle \beta_m \rangle = \frac{i g_{XX2}}{-\frac{\kappa_m}{2} + i(\omega_d - \omega_m)} \langle \alpha_{r2} \rangle + \frac{i g_{XX3}}{-\frac{\kappa_m}{2} + i(\omega_d - \omega_m)} \langle \alpha_{r3} \rangle
\end{align*}
and $n_m = |\langle \beta_m \rangle|^2$.

\section{4. Exciting magnons with the magnon driving line}
\label{magnon_num_top_CPW} 

Magnetic spin procession will induce an oscillating voltage in the magnon driving line with magnitude \cite{hou2021proposal}:
\begin{align*}
    V = b_{rf} \omega_m M_y,
\end{align*}
where $b_{rf} = {\mu_0}/{2 w}$ is the magnitude of the magnetic field experienced by magnet spins per unit current in the magnon driving line central wire of width $w$ when the spins are in close contact. $M_y$ is the total magnetic dipole moment oscillation amplitude along the width direction of the CPW. In terms of the number of spins $N$, the cone angle $\theta$ and Bohr magneton $\mu_B$, $M_y = \mu_B N \sin \theta \approx \mu_B N \theta$ for small $\theta$. 

%\noindent \gdf{I see how you derived this, but I'm wondering if it is better than using \\
%$M_y = \mu_B N \sin \theta \approx \mu_B N \theta$ \\
%and then below use \\
%$E=\mu_0 H N \mu_B(1-\cos \theta) \approx \mu_0 H N \mu_B\theta^2$\\
%and then your expression for $V$ could be derived as usual.  It seems more intuitive where the small angle approximation is coming from.  Your approach is correct as far as I can see (and possibly better because you make only one small angle approximation and not two) but it wasn't obvious.  I did a double-take when I first read it carefully.} 

Since the energy of magnetic excitation is $E=\mu_0 H N \mu_B(1-\cos \theta) \approx \mu_0 H N \mu_B\theta^2$, we have the induced voltage amplitude
\begin{align*}
    V &\approx b_{rf} \omega_m \mu_B N \theta\\
    &= b_{rf} \omega_m \mu_B N \sqrt{\frac{2E}{\mu_0 H N \mu_B}}.
\end{align*}
Using $\mu_0 H = \frac{\omega_m}{\gamma}$ (neglecting anisotropy), we have
\begin{align*}
    V = b_{rf} \sqrt{2\omega_m \mu_B N E \gamma}.
\end{align*}

The magnons lose energy through the magnon driving line because spin procession will induce a microwave voltage in the magnon driving line that will propagate in both directions away from the magnetic material (which we assume is the $5 \; \mu$m $\times \; 5 \; \mu$m $\times \; 1 \; \mu$m YIG). Note that the magnetic material's dimensions are much smaller than the MW wavelength of the magnon driving line. Since the spin procession will induce a potential difference of $V$, on each side of the magnetic material on the magnon driving line's central wire, the potential will be $V/2$ and $-V/2$ respectively. We can see this by applying the Faraday's law in Fig. \ref{flux_induced_emf}, which shows the top view of the YIG under the magnon driving line. Consider the rectangular path A$\rightarrow$B$\rightarrow$C$\rightarrow$D$\rightarrow$A. When the distance between D and A $l_{DA} >> 5 \; \mu$m, we can assume that this rectangle encloses all the oscillating stray field flux from the YIG, and the cyclic line integrals of oscillating electric field amplitude is
\begin{equation*}
    \oint_{\text{A} \rightarrow\text{B}\rightarrow\text{C}\rightarrow\text{D}\rightarrow\text{A}} \vec{E} \,d\vec{r} = V.
\end{equation*}

We can also assume $l_{DA}<<\lambda$ (the MW wavelength in the magnon driving line) and the inductance over the distance $l_{DA}$ is negligible, so that the cooper pair density in the Nb is adjusted with respect to time such that $\int_{\text{B}}^{\text{C}} \vec{E} \,d\vec{r} = \int_{\text{D}}^{\text{A}} \vec{E} \,d\vec{r} = 0$. So, by symmetry, $\int_{\text{B}}^{\text{A}} \vec{E} \,d\vec{r} = -V/2$ and $\int_{\text{C}}^{\text{D}} \vec{E} \,d\vec{r} = V/2$. Note that the voltage here is not a state function.

\begin{figure}
\includegraphics[width = 60mm]{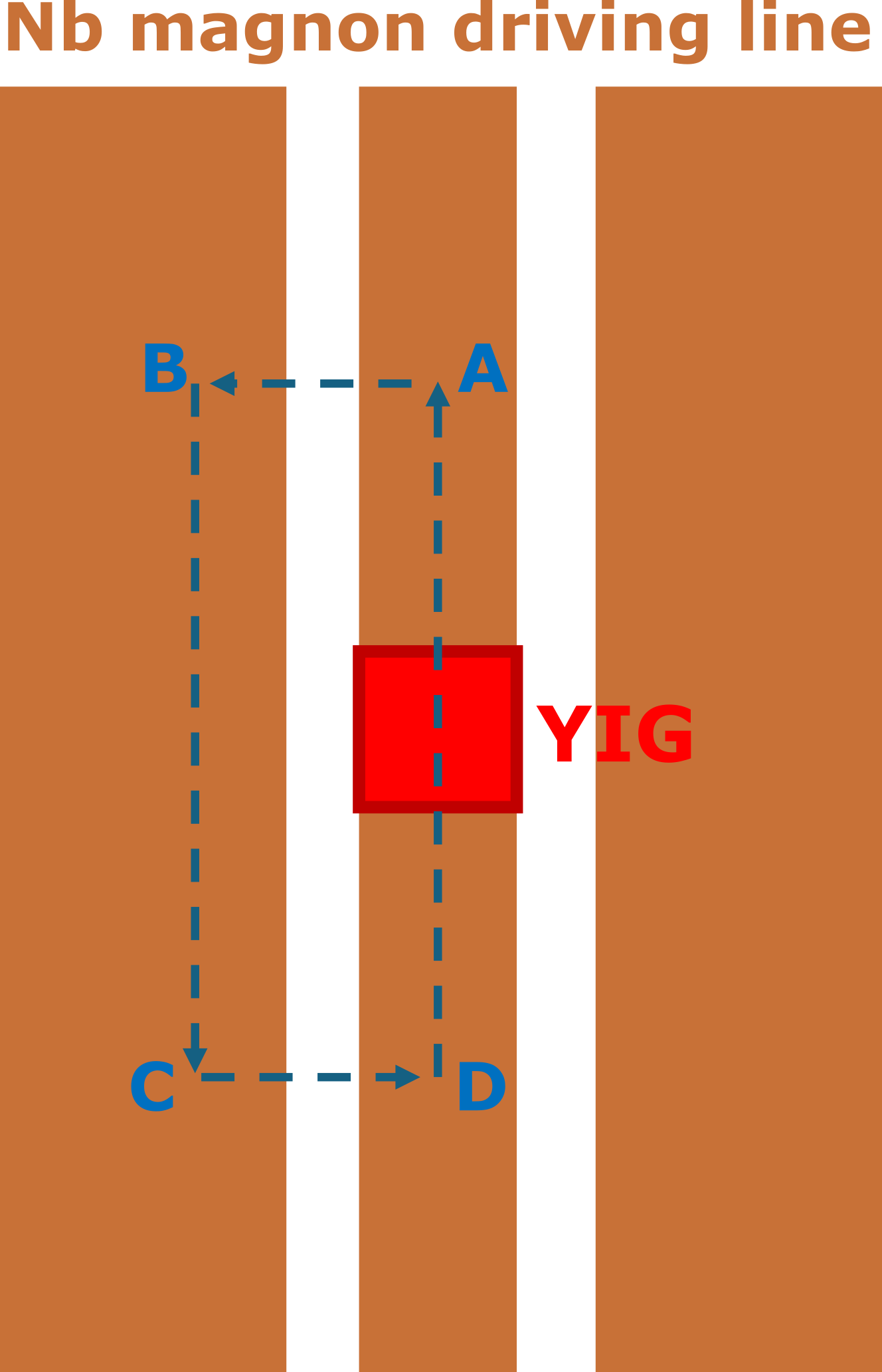}
\caption{Top view of the Nb magnon driving line and the YIG of the top CPW geometry.}
\label{flux_induced_emf}
\end{figure}

The power of magnons' external loss is thus
\begin{align*}
    P = 2\times \frac{(V/2)^2}{2Z_0} = \frac{{b_{rf}}^2 \omega_m \mu_B N E \gamma}{2 Z_0},
\end{align*}
where $Z_0 = 50 \; \Omega$ is the impedance of the coplanar waveguide.
So the external damping rate of the magnons is:
\begin{align*}
    \kappa_{m,ext} = \frac{P}{E} = \frac{{b_{rf}}^2 \omega_m \mu_B N \gamma}{2 Z_0}
\end{align*}
corresponding to a the coupling quality factor of:
\begin{align*}
    Q_{m,c} = \frac{\omega_m}{\kappa_{m,ext}} = \frac{2 Z_0}{{b_{rf}}^2 \mu_B N \gamma}.
\end{align*}

The saturation magnetization of YIG is $M_s = 140$ emu/cm$^3$ \cite{harris2012microwave}. With the $5 \; \mu$m wide CPW central wire and $5 \; \mu$m $\times \; 5 \; \mu$m $\times \; 1 \; \mu$m YIG, we find:
\begin{align*}
    Q_{m,c} = \frac{2 Z_0}{\gamma {b_{rf}}^2 \mu_B N} = \frac{2 Z_0}{\gamma {b_{rf}}^2 M_s V_{YIG}} = 10000,
\end{align*}
where $V_{YIG}$ is the volume of YIG.

Note that the above calculation does not take the anisotropy field into account. If we consider the anisotropy, the Q factor will be smaller by a factor of $\sqrt{\frac{H + M_{eff}}{H}}$ for the in plane static field in the thin film limit, where $M_{\text{eff}} = M_s - H_k$, and $H_k$ is the uniaxial anisotropy field.

Similar to exciting a superconducting resonator, when the MW frequency along the magnon driving line is $\omega_d$ and the power is $P$, the magnon oscillation amplitude is:
\begin{align*}
    \langle \beta_m \rangle = \sqrt{\frac{2P}{\hbar\omega_{d}}} \frac{\sqrt{\kappa_{m,ext}}}{-\kappa_m + 2i(\omega - \omega_{res})}
\end{align*}
and the average magnon number is:
\begin{align*}
    \langle n_m \rangle =  \left| \langle \beta_m \rangle \right|^2 = \frac{2P}{\hbar\omega_d} \frac{\kappa_{m,ext}}{\kappa_m^2 + 4(\omega - \omega_{res})^2}
\end{align*}
For example, when driving at the red sideband of YIG at 0~dBm, the average magnon number  $\langle n_m \rangle = 4.8 \times 10^7$. With a $5 \; \mu$m $\times \; 5 \; \mu$m $\times \; 1 \; \mu$m YIG, we have the total number of spins:
\begin{align*}
    N = \frac{M_s V_{YIG}}{\mu_B} = 3.8 \times 10^{11}
\end{align*}
So $\langle n_m \rangle << N$.

At such a large excitation power, the Kerr nonlinearity of magnons can be a problem for creating a resonator squeezed state.  In particular, if the magnon frequency shift is bigger than $\kappa_m$, it will complicate  driving on the red (blue) sidebands. The Kerr nonlinearity is represented by the term $K \hat{b}^{\dagger}\hat{b}\hat{b}^{\dagger}\hat{b}$ in the Hamiltonian, where $K \propto M_{eff}$ is the Kerr coefficient \cite{wang2016magnon}. In our top CPW geometry the Kerr coefficient from the shape anisotropy is $K = 0.01$~Hz. So, when $\langle n_m \rangle = 4.8 \times 10^7$, the frequency shift of magnons is 0.48~MHz, which is much smaller than $\kappa_m$. Although the Kerr nonlinearity will already be small, our scheme would work best if it is minimized. We can put strain on YIG to adjust the magnetocrystaline anisotropy $H_k$ and thus reduce the effective magnetization $M_{eff} = M_s - H_k$ \cite{wang2024giant}, where $M_s$ is YIG's saturation magnetization.

%At such a large excitation power, the Kerr nonlinearity of magnons can be a problem for creating the resonator squeezed state, since the frequency shift of magnons will prevent us from diving right at the red (blue) sidebands. To avoid this issue, we can put strain on YIG to adjust the magnetocrystaline anisotropy $H_k$ and thus reduce the effective magnetization $M_{eff} = M_s - H_k$ \cite{wang2024giant}, where $M_s$ is YIG's saturation magnetization. The Kerr nonlinearity will reduce with $M_{eff}$ since they are proportional \cite{wang2016magnon}. or we can use YIG sphere instead.

\section{5. Zero point current fluctuation ($I_{ZPF}$) of the quarter wavelength resonator fundamental mode}
\label{ZPF}
The impedance of the resonator (impedance of the CPW) is $Z_0 = \frac{1}{C_{CPW}v_{ph}}$, where $C_{CPW}$ is the capacitance per unit length and $v_{ph}$ is the photon propagation velocity in the CPW \cite{simons2004coplanar}. So the total capacitance of the CPW resonator is:
\begin{align*}
    C_{total} = C_{CPW} \times \frac{\lambda}{4} = \frac{1}{Z_0 v_{ph}} \times \frac{\lambda}{4} = \frac{1}{4 Z_0} \frac{2 \pi}{\omega_{r1}}.
\end{align*}
The total inductance of the CPW resonator is thus:
\begin{align*}
    L_{total} = C_{total} {Z_0}^2 = \frac{Z_0}{2} \frac{\pi}{\omega_{r1}}.
\end{align*}
Also, the zero point energy is:
\begin{align*}
    \frac{1}{2} \hbar \omega_{r1} = \frac{1}{2}L_{total}{I_{ZPF}}^2 = \frac{1}{2} \frac{Z_0}{2} \frac{\pi}{\omega_{r1}} {I_{ZPF}}^2.
\end{align*}
So the zero point current fluctuation at the current antinode $I_{ZPF} = \sqrt{\frac{2 \hbar}{\pi Z_0}}\omega_{r1}$, which is $\sqrt{\frac{4}{\pi}}$ larger than that of the lumped element resonator \cite{blais2021circuit} and is $\sqrt{2}$ times larger than that of the $\lambda/2$ resonator.

\section{6. External damping rate of microwave resonator modes}
\label{ext_damp_res}

The total capacitance of a quarter wavelength CPW resonator is $C_{total} = \frac{\pi}{2 Z_0 \omega_{r1}}$ (section 5). When the resonator's energy is $E$, the voltage oscillation amplitude at the charge antinode $V_{AC}$ satisfies:
\begin{align*}
    E = \frac{1}{4}C_{total}{V_{AC}}^2.
\end{align*}

When a coplanar waveguide is coupled to this resonator with coupling capacitance $C_c$, the induced oscillation charge amplitude on the CPW central wire is $Q_{AC} = C_c V_{AC}$ (assume the size of the coupling capacitor is much smaller than the wavelength of the driving MW). This oscillating charge will draw current from/to both directions along the resonator feedline. For each direction the oscillating current amplitude is $I_{AC} = \omega_{res} Q_{AC} / 2$. Where $\omega_{res}/2\pi =$ 0.5 GHz or 1.5 GHz or... are any of the harmonic modes of the $\lambda/4$ resonator. So the power lost from the resonator to the resonator feedline is:
\begin{align*}
    P = 2 \times \frac{1}{2} Z_0 I_{AC}^2 = \frac{1}{4} Z_0 \omega_{res}^2 C_c^2 V_{AC}^2 = \frac{Z_0 \omega_{res}^2 C_c^2 E}{C_{total}}.
\end{align*}
Thus the external damping rate of the oscillator is:
\begin{align*}
    \kappa_{res,ext} = \frac{P}{E} = \frac{Z_0 \omega_{res}^2 C_c^2}{C_{total}}
\end{align*}
corresponding to an coupling quality factor of:
\begin{align*}
    Q_{res,c} = \frac{\omega_{res}}{\kappa_{res,ext}} = \frac{C_{total}}{Z_0 \omega_{res} C_c^2}
\end{align*}

We see that $Q_{res,c}$ for these higher frequency modes will decrease with respect to $\omega_{res}$. In other words, the coupling capacitor acts as a high pass filter in the lumped element approximation. In the setup, we need to make the dimension of the coupling capacitor much smaller than the wavelength of the highest frequency mode for our experiment to make this approximation valid.

\section{7. XX coupling rate between the Kittel mode and the higher order resonator harmonic mode}

\label{XXharmonic}

When the external static filed is parallel to the inductor wire, the XX coupling strength between a lumped element resonator and magnetic material is \cite{Tabuchi2014, Hou2019, eichler2017electron} (If we consider the anisotropy, this value will be enhanced by a factor of $(\frac{H + M_{eff}}{H})^{1/4}$ for the in plane static field):
\begin{align*}
    g_{xx,LE} = g_e \mu_B b_{rf} \omega_{res} \sqrt{\frac{N}{8 \hbar Z_{LE}}},
\end{align*}
where $g_e$ is the electron Land\'e $g$ factor, $\mu_B$ is the Bohr magneton, $b_{rf} = {\mu_0}/{2 w}$ is the magnetic field magnitude at magnet spins per unit current in an inductor wire of width $w$ when the spins are in close contact, $N$ is the number of spins and $Z_{LE}=\sqrt{L_{LE}/C_{LE}}$ is the characteristic impedance of the lumped element resonator.

%where $N$ is the number of spins, $b_{rf} = {\mu_0}/{2 w}$ is the magnitude of the magnetic field experienced by magnet spins per unit current in an inductor wire of width $w$ when the spins are in close contact.  We have also used the electron Land\'e $g$ factor $g_e$, the Bohr magneton $\mu_B$, and the characteristic impedance of the lumped element resonator $Z_{LE}=\sqrt{L_{LE}/C_{LE}}$. (Copied from the strong coupling paper, do I need to paraphrase this?)

However, if we use the quarter wavelength reosonator instead (with the magnetic material at its current antinode), the AC current amplitude at the current antinode will be $\sqrt{\frac{4}{\pi}}$ higher than that of the lumped element resonator (for the same excitation energy, see section 5). So, for the fundamental mode of the $\lambda/4$ resonator:
\begin{align*}
    g_{XX1} = g_e \mu_B b_{rf} \omega_{r1} \sqrt{\frac{N}{2 \pi \hbar Z_{0}}}
\end{align*}
where $\omega_{r1}$ is the fundamental mode resonator frequency.

In our setup, the fundamental mode frequency is 0.5 GHz and we are considering the XX coupling between the magnetic material and higher order harmonic modes of the $\lambda/4$ resonator with for example $\omega_{r2}/2\pi = 19.5$ GHz. For the same excitation energy, the AC current amplitude at the current antinode of this resonator is $\sqrt{19.5/0.5}$ times smaller than that of the $\lambda/4$ resonator whose fundamental mode frequency is 19.5 GHz. Also, in the alternative setup our inductor wire is 45 degrees away from the static magnetic field, so the XX coupling between the magnetic material and the 19.5 GHz mode is:
\begin{align*}
    g_{XX2}/2\pi &= \sqrt{\frac{1}{2}} g_e \mu_B b_{rf} (19.5\text{GHz}) \sqrt{\frac{0.5\text{GHz}}{19.5\text{GHz}}} \sqrt{\frac{N}{2 \pi \hbar Z_0}}\\
    &= g_e \mu_B b_{rf} \sqrt{(0.5\text{GHz})(19.5\text{GHz})}\sqrt{\frac{N}{4 \pi \hbar Z_0}}
\end{align*}
With a $500 \; \mu$m $\times \; 5 \; \mu$m $\times \; 1 \; \mu$m \VTNCE, we have:
\begin{align*}
    N = \frac{M_s V_{VTCNE}}{\mu_B} = 2.2 \times 10^{12},
\end{align*}
where $V_{VTCNE}$ is the volume of \VTNCE. So $g_{XX2}/2\pi = 41.5$ MHz and $g_{XX3}/2\pi = 42.6$ MHz. 

If we use YIG instead, $M_s$ is around 17.6 times bigger, so $g_{XX2}/2\pi = 174$ MHz and $g_{XX3}/2\pi = 178$ MHz.

Note that we have assumed $\lambda/4 >> 500\;\mu$m for the calculation above. In reality, $\lambda/4$ is about 1.5 mm for the $20.5$ GHz mode, which is comparable to the length of the magnetic material. If we consider this, the resulting XX coupling rate will be slightly smaller.

%\bibliography{OM_BIB}

\end{widetext}

\end{document}